\documentclass{article}

\usepackage[preprint,nonatbib]{neurips_2023}

\usepackage[utf8]{inputenc} % allow utf-8 input
\usepackage[T1]{fontenc}    % use 8-bit T1 fonts
\usepackage{hyperref}       % hyperlinks
\usepackage{url}            % simple URL typesetting
\usepackage{booktabs}       % professional-quality tables
\usepackage{amsfonts}       % blackboard math symbols
\usepackage{nicefrac}       % compact symbols for 1/2, etc.
\usepackage{microtype}      % microtypography
\usepackage[dvipsnames]{xcolor}         % colors
\usepackage{tabu}
\usepackage{graphicx}
\usepackage[relative,overlay]{textpos}
\usepackage{amsmath}
\usepackage{fancyvrb}
\RequirePackage[skins]{tcolorbox}

\definecolor{cool}{RGB}{74,173,143}
\definecolor{cool-light}{RGB}{237,247,243}

\newtcolorbox{custombox}[1]{
	colback=cool-light,
	colframe=cool,
	left=1mm,
	right=1mm,
	top=1mm,
	bottom=1mm,
	fonttitle=\bfseries,
	arc=0mm,
	leftrule=1mm,
	rightrule=0mm,
	toprule=0mm,
	bottomrule=0mm,
	notitle,
	before=\par\smallskip\noindent,
	before upper={#1},
}

\newtoggle{anonreview}
\togglefalse{anonreview}

\newcommand{\toolname}{\iftoggle{anonreview}{ToolName}{CodeHelp}}

\title{%Less Data, No Problem: 
Efficient Classification of Student Help Requests in Programming Courses Using Large Language Models}

\author{%
  Jaromir Savelka\\
  Carnegie Mellon University\\
  Pittsburgh, Pennsylvania, USA \\
  \texttt{jsavelka@cs.cmu.edu} \\
  \And
  Paul Denny \\
  The University of Auckland \\
  Auckland, New Zealand \\
  \texttt{paul@cs.auckland.ac.nz} \\
  \And
  Mark Liffiton \\
  Illinois Wesleyan University \\
  Bloomington, Illinois, USA \\
  \texttt{mliffito@iwu.edu} \\
  \And
  Brad Sheese \\
  Illinois Wesleyan University \\
  Bloomington, Illinois, USA \\
  \texttt{bsheese@iwu.edu} \\
}

\begin{document}

\maketitle

\begin{abstract}
The accurate classification of student help requests with respect to the type of help being sought can enable the tailoring of effective responses.  Automatically classifying such requests is non-trivial, but large language models (LLMs) appear to offer an accessible, cost-effective solution. This study evaluates the performance of the GPT-3.5 and GPT-4 models for classifying help requests from students in an introductory programming class. In zero-shot trials, GPT-3.5 and GPT-4 exhibited comparable performance on most categories, while GPT-4 outperformed GPT-3.5 in classifying sub-categories for requests related to debugging. Fine-tuning the GPT-3.5 model improved its performance to such an extent that it approximated the accuracy and consistency across categories observed between two human raters. Overall, this study demonstrates the feasibility of using LLMs to enhance educational systems through the automated classification of student needs.

\end{abstract}

\section{Introduction}

The emergence of large language models (LLMs) has opened up new possibilities for enhancing educational tools and services.  In particular, one promising application of LLMs is providing personalized on-demand assistance at scale \cite{kasneci2023chatgpt,kazemitabaar2023studying}. This is especially valuable in courses with growing enrollments, such as introductory programming courses, where student-to-instructor ratios are large \cite{national2018assessing}.  When students seek help from an automated assistant, they may ask a wide range of different types of queries related to their programming assignments or projects.  The ability to classify these queries into distinct categories can have important educational implications, as evidenced by the related work (Section \ref{sec:related}). For example, if a student requests help implementing code directly related to an assignment, an appropriate response may be to restate the specifications more simply or ask the student to clarify what is unclear to them.  On the other hand, if a student requests assistance in debugging code, then a targeted hint about resolving the bug may be a useful response.  Moreover, identifying the types of queries that students tend to ask most frequently can be valuable feedback for instructors and researchers. 

Classifying student queries into suitable categories is difficult and time consuming as queries can differ in subtle ways and require expert knowledge to assess reliably.  In addition, automatic classification is necessary for integration into a tool, but expensive because it typically requires large amounts of expert-labeled data for training classifiers.  Given the recent successes of LLMs in computing education \cite{prather2023robots,leinonen2023comparing,sarsa2022automatic,savelka2023thrilled}, we explore the viability of using GPT-3.5 and GPT-4 to automatically classify student help requests when there is either little or no labeled training data available. Specifically, our research questions were as follows:

\begin{itemize}
    \item[(RQ1)] How accurately can GPT-3.5 and GPT-4 perform zero-shot classification of student help requests based on the coding instructions originally designed for human raters?
    \item[(RQ2)] To what extent can classification performance be improved by fine-tuning using a limited amount of data?
\end{itemize}

\section{Related Work}
\label{sec:related}
% Automatic classification
Researchers have long been interested in automatically categorizing student requests for online help and educational forum posts. Gao et al. used a gradient boosting framework to classify student help requests based on the sufficiency of information provided (e.g., \emph{useless}, \emph{sufficient}, or having a \emph{copied error}) \cite{gao2021automatically}. Similarly, \v{S}v\'abensk\'y et al. used several traditional ML algorithms (e.g., random forest or linear regression) to classify student posts according to their urgency on an ordinal scale (e.g., \emph{not actionable}, \emph{extremely urgent}) \cite{vsvabensky2023towards}. Xu and Lynch utilized a combination of a convolutional neural network and long short term memory model (CNN-LSTM), and Bi-directional LSTM (BiLSTM) to automatically classify MOOC discussion posts as to whether they were \emph{seeking help}, and to identify what kind of question was being asked (\emph{course content}, \emph{technique}, or \emph{course logistic}) \cite{xu2018you}. Sha et al. compared several traditional ML algorithms (e.g., random forest) to deep learning algorithms~(e.g., CNN, LSTM) on classifying student forum posts from two datasets---the Stanford MOOC posts dataset \cite{agrawal2015} encoding, e.g., \emph{urgency} or \emph{sentiment} of the posts, and their own dataset with posts labeled as \emph{content} and \emph{process} \cite{sha2022latest}. Onan and To\c{c}o\u{g}lu utilized clustering (unsupervised -- no training data required) to assign student questions to topic categories\cite{onan2021weighted}.

A number of studies have demonstrated the value of classifying student help requests and forum posts by manually categorizing them into schemes based on the nature of the questions. For example, Gao et al. analyzed the proportion and evolution over time of student request types, dividing them into eight categories related to, e.g., \emph{general debugging and addressing issues} or \emph{implementation and understanding} \cite{gao2022you,gao2022admitting}. Vellukunnel et al. analyzed discussion forum posts, distinguishing student posts where students did not demonstrate effort (\emph{active}) from posts that did showed effort to solve a problem~(\emph{constructive}) \cite{vellukunnel2017deconstructing}. 
To date, classification has required time-intensive manual coding by researchers. However, LLMs have the potential to enable faster, cheaper, and more consistent analysis of patterns and trends in student queries as evidenced by work in other domains \cite{savelka2023unlocking,savelka2023unreasonable}.

Several studies have investigated unproductive help-seeking behaviors in tutoring systems, such as \emph{help abuse} and \emph{try-step abuse}, in both general tutoring \cite{roll2006help,roll2011improving} and programming contexts \cite{marwan2020unproductive}. 
However, there is limited research on leveraging help request categorization to improve interactions in programming assistance chatbots. To our knowledge, Carreira et al. has developed the only programming chatbot (Pyo) that utilizes predefined categories for student questions like (\emph{exercise assistance}, \emph{error guidance}, \emph{concept definitions}) \cite{carreira_pyo_2022}. Although programming chatbots are an active research area, most do not distinguish between different student question types (e.g., Python-bot \cite{okonkwo_python-bot_2021}, RevBot \cite{okonkwo_revision-bot_2022}, Duckbot \cite{rutgers_duckbot_nodate} and others \cite{konecki_intelligent_2015,walden_chatbot_2022}). Existing systems do not tailor responses based on the intent behind students' inquiries. Categorizing questions allows personalized interactions that target the specific help students request. This study provides an initial step in demonstrating the feasibility of using query categorization to improve programming chatbots.

\section{Dataset}
One of the authors of this paper developed CodeHelp, an automated assistant that responds to semi-structured student queries in programming and CS courses \cite{liffiton2023codehelp}.
CodeHelp uses LLMs to generate responses to requests posed in natural language. Students request help via a form with separate inputs for the programming language they are using, a snippet of relevant code, an error message if they have one, and a question or description of the issue they are facing. Responses are generated by a series of prompts to LLMs.
One prompt checks whether the student's inputs are sufficient to be able to provide them with effective help, and if additional information is needed, it generates a request for clarification that is presented to the student. Another prompt, run concurrently, combines the student's inputs with instructions to provide guidance and explanations, along with class-specific context provided by the instructor, and its completion is used as the main response for the student.

We deployed \toolname{} in two sections of an undergraduate introductory and data-science course, totalling 52 students, taught by an author of this paper in the Spring semester of 2023. During the course, students submitted 2,082 unique queries requesting help. As reported in \cite{sheese2023patterns}, the queries were independently coded by two of the authors into the following categories: 

\begin{enumerate}
\item \emph{Debugging}: Queries seeking help to resolve errors in code; sub-categorized into queries that included: a) the error (dr); b) the desired outcome (dx); or c) both (drx). 
\item \emph{Implementation} (i): Queries about implementing code to solve specific assignment problems.
\item \emph{Understanding} (u): Queries focused on gaining an understanding of programming concepts.
\item \emph{Nothing} (n): Queries that provided no error or meaningful issue.
\end{enumerate}

Human raters showed substantial reliability for ratings of all categories and sub-categories ($\kappa=.75$). Overall reliability was even higher ($\kappa=.83$) when Debugging sub-categories were collapsed into a single Debugging category \cite{o2020intercoder}. For the current research, if there was disagreement between human raters, we used the rating from the more experienced rater as the "gold-label" classification.

\begin{figure}
\flushright
\includegraphics[width=.5\textwidth]{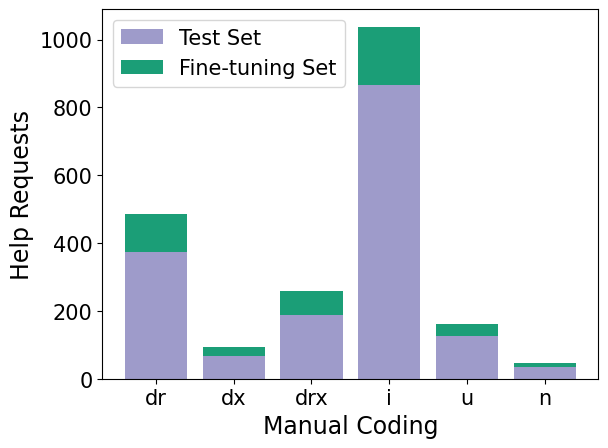}
\begin{textblock*}{6.8cm}(0.0cm,-5.20cm)
\begin{custombox}
 
\textbf{Language:}\\ Python \\
\textbf{Code:}\\ \texttt{major\_series.sort\_index['MUSIC']} \\
\textbf{Error:}\\ \texttt{'method' object is not subscriptable} \\
\textbf{Issue:}\\ I'm trying to index and find the value associated with the index label ('music') but it keeps saying this. How do I make it at least not give me an error?
\end{custombox}
\end{textblock*}
\caption{An example student help request is shown on the left. Counts of help requests by coding category and by data set split are shown on the right (the request codes are explained in the text).}
\label{fig:dataset}
\end{figure}

We divided the data set into a fine-tuning set and a test set. The fine-tuning set was used to fine-tune an LLM.
The split was performed on the basis of students, i.e., all the help requests submitted by a specific student were included in the same set. We randomly selected 10 students and included their requests in the fine-tuning set. Considering the number of requests submitted by each student, we made sure that two of the selected students were from the lowest quartile, six from second and third quartiles, and two from fourth quartile. Out of the 2,082 help requests, 423 were selected for the fine-tuning set, and the remaining 1,659 were included in the test set (see Figure \ref{fig:dataset}).

\section{Experiments}
\label{sec:experiments}

\paragraph{Models}
The original GPT model's core capability is \emph{fine-tuning} on a downstream task \cite{radford2018improving}. 
The GPT-2 model displays remarkable \emph{multi-task learning} capabilities~\cite{radford2019language}. 
The main focus of \cite{brown2020language} was to study the dependence of performance on model size where eight differently sized models were trained---the largest of the models is commonly referred to as GPT-3 (175 billion parameters). The interesting property of these very large models is that they appear to be very strong \emph{zero- and few-shot learners}~\cite{brown2020language}. The work of \cite{ouyangtraining} focused on the \emph{alignment problem}, demonstrating the apparent usefulness of fine-tuning the LLMs to follow instructions~(RLHF). In this paper, we evaluated \texttt{gpt-3.5-turbo-0613} and \texttt{gpt-4-0613} \cite{openai2023gpt}---some of the recently released GPT models.

\paragraph{Baselines} BERT (bidirectional encoder representation from transformers) \cite{devlin2018bert,vaswani2017attention} has gained immense popularity. A large number of models using similar architectures have been proposed~\cite{Lan2019,raffel2020exploring}, including RoBERTa (robustly optimized BERT pretraining approach) \cite{Liu2019}. A base model of RoBERTA (125 million parameters) is used as a baseline in the current study. A \emph{random forest} \cite{ho1995random} is an ensemble classifier that fits a number of decision trees on sub-samples of the data set. We included a random forest model in our experiments so that we could compare the GPT models to a well-regarded traditional ML technique.

\paragraph{Experimental Design}
In the zero-shot settings, we submit the requests from the test set one by one using the \texttt{openai} Python library\footnote{OpenAI Python Library. Available at: \url{https://pypi.org/project/openai/0.28.0/} [2023-09-17]} which is a wrapper for the OpenAI's REST API.\footnote{We set the \texttt{temperature} 0.0 (no randomness), \texttt{max\_tokens} to 10 (a response is a single label consisting of 1--3 letters), \texttt{top\_p} to 1 (recommended  when \texttt{temperature} is set to 0.0), and both \texttt{frequency\_penalty} and \texttt{presence\_penalty} to 0 (no penalty to repetitions or to tokens appearing multiple times in the output).} In our experiments we did not encounter any issues stemming from the models' prompt length limitations. Consequently, we neither adapted nor explored any measures to mitigate prompt length limitation issues. We included the coding instructions originally designed for human raters in the system part of the prompt (Appendix \ref{app:sys_prompt}) and a student help request in the user message (Appendix \ref{app:user_prompt}),  which were then combined into the prompt directly provided to the LLM to generate the completion. Each prompt completion (response) returned a predicted label, which we then compared to the gold-label~(i.e., the human-assigned category). We fine-tuned the \texttt{gpt-3.5-turbo-0613} model on 50, 100, 200 and all 423 student help requests included in the fine-tuning set. Hence, we could observe the effects of fine-tuning on progressively larger data sets. Each data point was structured following the exact same format of the system part of the prompt and the user message described above. All of the models were fine-tuned for 3 epochs.
To evaluate the performance of the models, we used Precision~($P$), Recall ($R$), and $F_1$-measure.  

\section{Results}
\begin{figure}
    \centering
    \begin{textblock*}{.375\textwidth}(-0.4cm,0cm)
        \includegraphics[width=\textwidth]{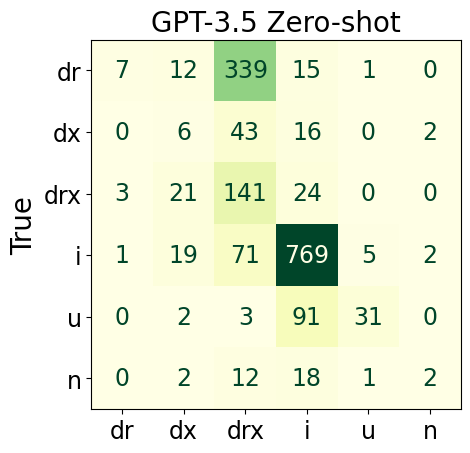}
    \end{textblock*}
    \includegraphics[width=.31\textwidth]{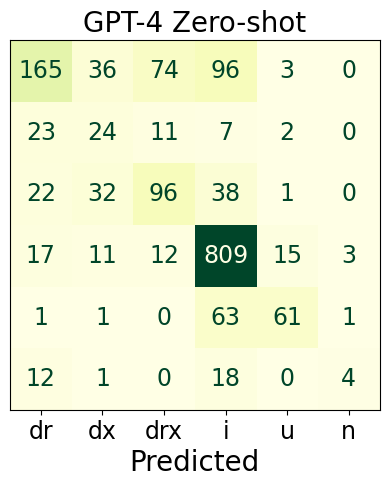}
    \begin{textblock*}{.31\textwidth}(9.12cm,-5.42cm)
        \includegraphics[width=\textwidth]{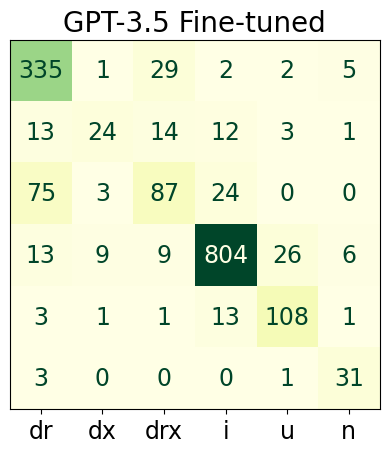}
    \end{textblock*}
    \caption{Confusion matrices of GPT-3.5 and GPT-4 classification output in zero-shot settings and when fine-tuned on 423 student help requests (refer to Table \ref{tab:results} for codes).}
    \label{fig:results-cm}
\end{figure}

Table \ref{tab:results} shows per class metrics as well as their overall weighted averages. The performance on the four main categories appeared to be similar for both, \texttt{gpt-3.5-turbo-0613} and \texttt{gpt-4-0613}, in the zero-shot settings, achieving the overall $F_1$ score of .81. There was a noticeable difference when it came to handling the \emph{Debugging} sub-categories. When these were considered, the GPT-4 model achieved overall $F_1=.68$ while the $F_1$ score of the GPT-3.5 model dropped to $.53$. Closer examination shows that, the drop was explained by the 339 \emph{Debugging -- error (dr)} help requests that were predicted as \emph{Debugging -- error \& outcome (drx)} by the GPT-3.5 model (Figure \ref{fig:results-cm}). This was also reflected in the $\kappa$ agreement scores with the manually assigned codes. Both the models achieved similar agreement with the gold-labels on the four main categories ($\kappa=.69$ for GPT-3.5, $\kappa=.67$ for GPT-4). When the \emph{Debugging} sub-categories were considered, the GPT-4 model ($\kappa=.52$) clearly outperformed the GPT-3.5 model ($\kappa=.36$). 

\begin{table}
    \centering
    \caption{Evaluation metrics examining the performance of GPT-3.5 and GPT-4 in zero-shot settings and when fine-tuned on 423 student help requests.}
    \setlength{\tabcolsep}{5.5pt}
    \label{tab:results}
    \begin{tabu}{lr|ccccccccc|ccc}
    \toprule
                                      & & &\multicolumn{7}{c}{ZERO-SHOT}& &\multicolumn{3}{c}{FINE-TUNED}\\
                                      & & &\multicolumn{3}{c}{GPT-3.5}& &\multicolumn{3}{c}{GPT-4}& &\multicolumn{3}{c}{GPT-3.5}\\
    Query Category                    &Count& & P & R & F$_1$ & & P & R & F$_1$ & & P & R & F$_1$\\
    \midrule
    Debugging                         &  630& &.84&.91&.87    & &.90&.77&.83    & &.94&.92&.93\\
    \rowfont{\color{gray}}           
    \hspace{.2cm} (error) -- dr            &  374& &.64&.02&.04    & &.69&.44&.54    & &.76&.90&.82\\
    \rowfont{\color{gray}}
    \hspace{.2cm} (outcome) -- dx           &   67& &.10&.09&.09    & &.23&.36&.28    & &.63&.36&.46\\      
    \rowfont{\color{gray}}           
    \hspace{.2cm} (error \& outcome) -- drx  &  189& &.23&.75&.35    & &.50&.51&.50    & &.62&.46&.53\\
    Implementation -- i                    &  867& &.82&.89&.85    & &.78&.93&.85    & &.94&.93&.93\\
    Understanding -- u                     &  127& &.82&.24&.38    & &.74&.48&.58    & &.77&.85&.81\\
    Nothing -- n                           &   35& &.33&.06&.10    & &.50&.11&.19    & &.70&.89&.78\\
    \midrule
    \rowfont{\bf}
    \bf Overall                       & 1659& &.82&.83&.81    & &.82&.82&.81    & &.92&.92&.92\\
    \rowfont{\bf \color{gray}}
    \hspace{.2cm}\bf (debugging types)& 1659& &.67&.58&.53    & &.70&.70&.68    & &.83&.84&.83\\
    \bottomrule
    \end{tabu}
\end{table}

Fine-tuning the \texttt{gpt-3.5-turbo-0613} model substantially improved performance. GPT-3.5 fine-tuned on all the 423 data points from the fine-tuning set achieved the overall $F_1$ scores of .92 (top-level categories) and .83 (with \emph{Debugging} sub-categories). The agreement of the fine-tuned model with the gold standard matched the agreement between the two human raters---$\kappa=.75$ ($\kappa=.75$ human) with \emph{Debugging} sub-categories and $\kappa=.86$ ($\kappa=.83$ human) when \emph{Debugging} sub-categories were collapsed. Figure \ref{fig:results-cm} provides detailed insight into the differences in handling the student help request classification task between the GPT-3.5 (zero-shot and fine-tuned) and GPT-4 (zero-shot).

\begin{figure}
    \centering
    \includegraphics[width=.42\textwidth]{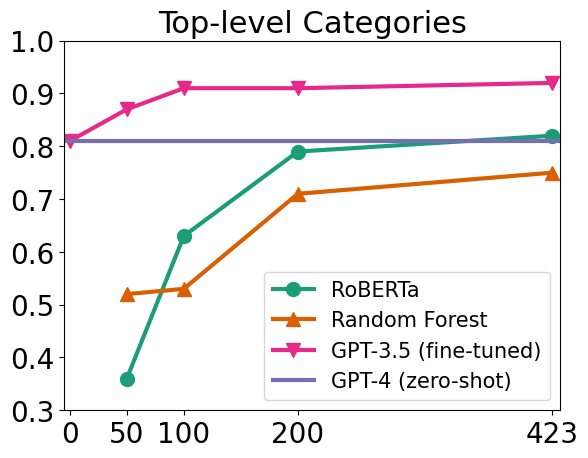}
    \hspace{0.2cm}
    \includegraphics[width=.42\textwidth]{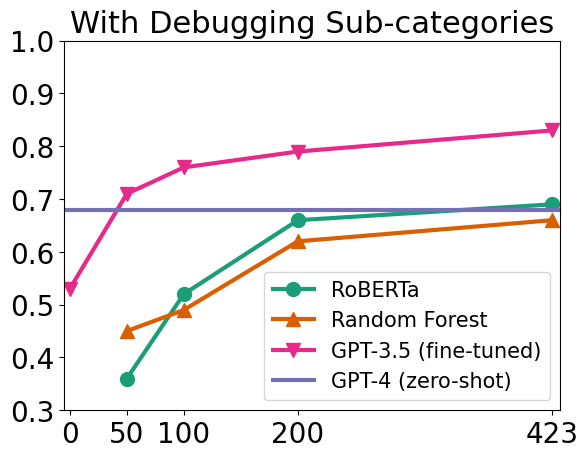}
    \caption{The comparison of GPT-4 and GPT-3.5 performance compared to random forest and RoBERTa base when trained/fine-tuned on progressively larder pool of data points up to 423.}
    \label{fig:baseline-cmp}
\end{figure}

Figure \ref{fig:baseline-cmp} demonstrates the key benefit of performing the classification with LLMs, such as GPT-3.5 or GPT-4, as compared to traditional ML algorithms or smaller LLMs. The LLMs performed reasonably even if no or very little ($n<100$) labeled data were available. A smaller LLM (RoBERTa base) required several hundred labeled data points to match the zero-shot performance of GPT-4, while a traditional ML algorithm such as random forest required even more labeled data (consistent with findings in other domains \cite{savelka2023unlocking,savelka2023unreasonable}). A small amount of labeled data ($n\simeq100$) was sufficient for the fine-tuned GPT-3.5 to perform comparably to humans on the easier task of labeling the four top-level categories. While it was also possible to match human performance on the more challenging task with the \emph{Debugging} sub-categories, a larger amount of labeled data was required ($n\simeq400$).

\section{Discussion}
Our results suggest that LLMs can perform classification tasks like ours on student help requests both accurately and inexpensively.
Compared to human raters, LLMs reach similar levels of performance at a very \emph{small fraction of the cost}, with much higher speed, low setup complexity, and greater flexibility to adapt to new or modified labeling schemes and educational contexts.
This can enable novel features in automated assistance systems such as \toolname{}. As to the cost, the fine-tuning of the \texttt{gpt-3.5-turbo-0613} on the 423 requests was performed over 3 epochs (1,269 steps). The overall number of submitted tokens was 1,003,722. At the time, the cost of fine-tuning the model was set to \$0.008/1K tokens.\footnote{OpenAI: Pricing. Available at: \url{https://openai.com/pricing} [Accessed 2023-09-17]} Hence, the overall cost of the procedure was \$8.03. For fine-tuning on 50 requests (117,609 tokens), the price was \$0.94. The current cost of using a fine-tuned GPT-3.5 model was \$0.012/1K for input tokens and \$0.016/1K for completions. The employment of the fine-tuned model as a classification component in \toolname{} over the Spring'23 semester would have amounted to less than \$30 additional cost.\footnote{2,591 submitted requests (not de-duplicated) with length of 1,000 input tokens and 10 for completions.} Using the general (not fine-tuned) GPT-3.5 model would cost less than \$4 while GPT-4 would cost roughly \$80.

By automatically classifying student requests into types, an LLM-powered system can provide instructors with rich, real-time aggregated information about their students' questions and help-seeking behaviors both across a class and for individual students.
This could allow an instructor to, for example, identify a shift in query types within a class that could suggest an increased difficulty in a module. Similarly, they could observe a heavy reliance on one type of query by an individual student, such as a student only ever asking \emph{Debugging} questions without providing an error or incorrect outcome. This could trigger an intervention to identify the cause and help the student improve their approach. The system itself could also utilize the classification as part of its operation to improve its responses.
For example, it could use the classification of a query to choose among different specialized prompts when generating a main response.
The user interface could automatically request additional information from the user when certain query types are recognized.
Students often do not know how to communicate effectively about technical subjects, and automated classification of their requests can play an important role in guiding them to more effective requests.

Using LLMs as a service to perform classification tasks is more accessible than using other ML-powered methods.
LLMs are hosted and available via APIs, requiring little to no local infrastructure and relatively little technical expertise.
The available models perform reasonably well with no labeled data. In our context, a small amount of labeled data to fine-tune a model yielded performance similar to that of human classifiers.
The fine-tuning is performed via an API as well, and it is fast and inexpensive.
This all suggests both an ease of integration with existing systems and a low barrier to experimenting with many different labeling schemes.
This allows tailoring a system to specific educational contexts as well as rapid iterative improvement of existing schemes.

\paragraph{Limitations} This study is an initial exploration rather than a comprehensive benchmark. We did not seek to maximally optimize model performance and do not claim that our results allow models to exhibit their best performance. Thus, this study should not be viewed as precisely measuring model capabilities, but rather hinting at the potential of LLMs in zero-shot settings or with minimal fine-tuning. There are several potential avenues to explore with regards to improving performance: prompt instructions for classification included an unaltered copy of the coding instructions developed for human raters. The prompt also included instructions to prevent the model from explaining its predictions, but generating an explanation followed by a prediction could lead to improvements \cite{wei2022chain}. More thorough experimentation with hyper-parameters could yield improved performance across all the studied models. It is not clear to what degree our findings would generalize to student queries from other courses, or other query classification schemes, or to non-English speaking courses. 

\section{Conclusions and Future Work}
We explored the use of LLMs for the classification of student help requests in introductory programming classes. We found that GPT-3.5 and GPT-4 models achieved reasonable accuracy in a zero-shot setting. Our results also showed that fine-tuning the GPT-3.5 model on a small amount of labeled data greatly improved its performance, reaching human-level accuracy. Our findings have important implications for personalized and scalable assistance in education. Automated systems that accurately classify student queries can provide tailored and effective responses to students and deliver insights to educators about how students are interacting with such tools. 

For future work, it would be valuable to explore the generalizability of our methods to other disciplines and model architectures. Additionally, further research can investigate the impact of different prompt instructions and hyper-parameters on the performance of LLMs for student query classification. Furthermore, it would be worthwhile to study the potential of fine-tuned LLMs in improving student interactions with assisting chatbots.

\bibliography{main} 
\bibliographystyle{plain}

\newpage
\appendix
\section{GPT Prompts}
\label{app:prompts}

\subsection{System Part of the Prompt}
\label{app:sys_prompt}
The system part of the prompt is typically used to steer the GPT dialogue-focused models towards performing the desired task. We introduced only minimal changes to the coding instruction to ensure close mapping between the original task performed by human raters and the task performed by GPT-3.5 and GPT-4 automatically. Below is an excerpt from the system part of the prompt used in our experiment. The gray \texttt{[...]} tokens indicate a part has been left out for brevity of the presentation.

\begin{figure}[h]
\scriptsize
\begin{Verbatim}[frame=single,commandchars=\\\{\}]
You are an educational assistant bot focused on analysis of student help requests in introductory 
programming courses. Given a help request you assign it with one of the below defined codes.

CODES
dr - Debugging (with error)
dx - Debugging (with expected outcome)
drx - Debugging (with error and expected outcome)
u - Understanding
i - Implementation / How to
n - Nothing

DEBUGGING REQUESTS
Cases where students are looking for help to solve specific errors and faults in their code. Must include 
some description of or pointer to an error -or- an indication of what it was supposed to do (even if no 
error is provided).

Typical requests in this category:
- Why doesn't/can't/isn't the code X
- The code doesn't do X
- It is supposed to X but it is doing Y
- I am trying to do X but my code does Y
- Error submitted with no description or context
- Error submitted with desired outcome described

Sub-classifications (either or both may be true, must be at least one):
dx) Does the query include an indication of the problem or incorrect outcome? (Error message or 
    description of what it does that is wrong.)
dr) Does the query include an indication of what the code is supposed to do?
drx) Does the query include both an indication of the problem and what the code is supposed to do?

GUIDANCE ON CODE IMPLEMENTATION 
Queries about implementing code or functions to solve specific assignment problems. Must include code 
and/or reference assignment instructions. High issue-instructions equivalence indicates that the student 
is likely asking for help with a specific assignment problem.

Typical requests in this category:
- How do I create X
- How do I write a function that X
- How do I get it to X
- I want to X
- No request, just restates course instructions

DEVELOPING UNDERSTANDING
Requests centered around gaining an understanding of programming concepts, algorithms, data structures,
language and library features but not obviously asking how to complete a given assignment problem. Must 
not include code or reference assignment instructions.

\textcolor{gray}{[...]}

INPUT
As input you will receive four elements:
Issue - a string describing the issue if any
Code - a string containing the code submitted by the student if any
Error - a string containing the error message if any
Issue-instructions equivalence - a percentage indicating how much the issue matches the assignment 
instructions

OUTPUT
As a response to the provided help request return one of the codes. Do not provide any explanations.

EXAMPLE OUTPUT 1
i

EXAMPLE OUTPUT 2
drx \textcolor{gray}{[...]}
\end{Verbatim}
% \begin{textblock*}{3.4cm}(6.3cm,-2.7cm)
% \circled{1}
% \end{textblock*}
% \begin{textblock*}{3.4cm}(1.5cm,-1.95cm)
% \circled{2}
% \end{textblock*}
% \begin{textblock*}{3.4cm}(0.1cm,-1.1cm)
% \circled{3}
% \end{textblock*}
%\caption{Selected excerpts from the system part of the prompt are shown at the top. The user message template is shown in the middle. The bottom section has the template filled in with a request.}
\label{fig:system-prompt}
\end{figure}

\subsection{User Message Template}
\label{app:user_prompt}
A student help request is provided via a user message. The green tokens with curly braces are replaced with the actual data. The filled-in user message is combined with the system part of the prompt and submitted as a prompt to an LLM, which is expected to generate a completion.

\begin{figure}[h]
\scriptsize
\begin{Verbatim}[frame=single,commandchars=\\\{\}]
Student Message: \textcolor{PineGreen}{\bf\string{\string{issue_description\string}\string}}

Student Code: \textcolor{PineGreen}{\bf\string{\string{code\string}\string}}

Student Error: \textcolor{PineGreen}{\bf\string{\string{error\string}\string}}

The \textcolor{PineGreen}{\bf\string{\string{issue_eq\string}\string}}% of the described student message is copied from course assignment instructions or code.
\end{Verbatim}
% \begin{textblock*}{3.4cm}(6.3cm,-2.7cm)
% \circled{1}
% \end{textblock*}
% \begin{textblock*}{3.4cm}(1.5cm,-1.95cm)
% \circled{2}
% \end{textblock*}
% \begin{textblock*}{3.4cm}(0.1cm,-1.1cm)
% \circled{3}
% \end{textblock*}
%\caption{Selected excerpts from the system part of the prompt are shown at the top. The user message template is shown in the middle. The bottom section has the template filled in with a request.}
\label{fig:user-prompt-template}
\end{figure}

\subsection{User Message Example}
An example of filled-in user message template is shown below.

\begin{figure}[h]
\scriptsize
\begin{Verbatim}[frame=single,commandchars=\\\{\}]
Student Message:
I’m trying to index and find the value associated with the index label (’music’) but it keeps saying 
this. How do I make it at least not give me an error?

Student Code:
major_series.sort_index[’MUSIC’]

Student Error:
'method' object is not subscriptable

The \textcolor{PineGreen}12.65% of the described student message is copied from course assignment instructions or code.
\end{Verbatim}
% \begin{textblock*}{3.4cm}(6.3cm,-2.7cm)
% \circled{1}
% \end{textblock*}
% \begin{textblock*}{3.4cm}(1.5cm,-1.95cm)
% \circled{2}
% \end{textblock*}
% \begin{textblock*}{3.4cm}(0.1cm,-1.1cm)
% \circled{3}
% \end{textblock*}
%\caption{Selected excerpts from the system part of the prompt are shown at the top. The user message template is shown in the middle. The bottom section has the template filled in with a request.}
\label{fig:user-prompt-example}
\end{figure}

\end{document}